\input{aipcheck.tex}

\edef\optionlist{%
   \variorefoptionifavailable        
   draft,%
   \psnfssproblemoption              
   tnotealph}

\documentclass[\optionlist]{aipproc}

\layoutstyle{8x11single}

\received{x}
\accepted{x}

\usepackage{ttct0001}

\usepackage{shortvrb}
\MakeShortVerb\|

\makeatletter
   \def\@oddfoot{\reset@font
                 \copyright{} 2004 AIP
                 \hfil\@title
                 \hfil\@date\hfil\thepage}
\makeatother

\begin{document}

\author{L\'aszl\'o \'{A}. Gergely}{
  address={Departments of Theoretical and Experimental Physics, University of Szeged, D\'om t\'er 
9, Szeged 6720, Hungary},
  email={gergely@physx.u-szeged.hu},
}

\author{Zolt\'an Keresztes}{
  address={Departments of Theoretical and Experimental Physics, University of Szeged, D\'om t\'er 
9, Szeged 6720, Hungary},
  email={zkeresztes@titan.physx.u-szeged.hu},
}

\author{Gyula M. Szab\'{o}}{
  address={Departments of Theoretical and Experimental Physics, University of Szeged, D\'om t\'er 
9, Szeged 6720, Hungary},
  email={szgy@titan.physx.u-szeged.hu},  
}

\title{Cosmological tests of generalized RS brane-worlds with Weyl fluid}

\date{\today}

\keywords{brane-worlds, Weyl-fluid, supernovae}
\classification{43.35.Ei, 78.60.Mq}

\begin{abstract}
A class of generalized Randall-Sundrum type II (RS) brane-world models with Weyl fluid are confronted with the 
Gold supernovae data set and BBN constraints. We consider three models with different evolutionary history of the Weyl 
fluid, characterized by the parameter $\alpha$. For $\alpha =0$ the Weyl curvature of the bulk appears as dark radiation 
on the brane, while for $\alpha =~2$ and $~3$ the brane radiates, leaving a Weyl fluid on the brane with energy density 
decreasing slower than that of (dark) matter. 
In each case the contribution $\Omega_d$ of the Weyl fluid represents but a few percent of the energy content of 
the Universe. 
All models fit reasonably well the Gold2006 data. The best fit model for $\alpha =0$ is for $\Omega_d=0.04$. 
In order to obey BBN constraints in this model however, the brane had to radiate at earlier times. 
\end{abstract}

\maketitle


\bigskip

\section{Introduction}

The $\Lambda $CDM model according to which our Universe is a Friedmann-Lemaitre-Robertson-Walker (FLRW) space-time 
with flat spatial sections containing approximately 3\% baryonic matter, 24\% cold dark matter, the rest being given by the 
contribution of a cosmological constant $\Lambda$ seems to be in excellent agreement with current observational data.
As the dark sector (dark matter and dark energy - a generalization of the vacuum expectation energy 
represented by the cosmological constant) remains unknown, alternate gravitational theories have been advanced.

The string-theory motivated brane-world models contain our observable Universe as a time-evolving 3-dimensional brane 
embedded in a 5-dimensional bulk. Standard model fields act on the brane, but gravity is allowed to leak into the fifth 
dimension \cite{RS2}, where non-standard model fields could also exist (for a review see \cite{MaartensLR}). 
The projection of the 5-dimensional Einstein equation onto the brane generates an effective Einstein equation with new 
source terms as compared to general relativity \cite{SMS}, \cite{Decomp}. 

Among them, the energy-momentum squared source term modifies early cosmology \cite{BDEL} and becomes important during the final 
stages of gravitational collapse \cite{collapse}. It behaves as the dominant source term before the Bing Bang 
Nucleosynthesis (BBN). This quadratic source term is proportional to $1/\lambda$. The value of the brane tension $\lambda$ 
is constrained by the deviation from the gravitational Newton-law still compatible with nowadays rigurous 
experiments \cite{tabletop}-\cite{GK}. 
The emerging high value of $\lambda$ and the fast decrease of the square of the energy density of matter implies that in 
a cosmological context this source term can be safely ignored at present time.   

The Weyl curvature of the bulk gives rise to a non-local bulk effect on the brane, appearing as fluid on the brane 
(the Weyl fluid). In the simplest case, when the bulk is static, the Weyl fluid is a radiation field (dark radiation). 
This situation represents an equilibrium configuration, without any energy exchange between the brane and the bulk. 
BBN constraints the amounts of the energy density of the dark radiation as 
$-1.02\times 10^{-4}\leq \Omega _{d}\leq 2.62\times 10^{-5}$ \cite{BBN}.
(Here $\Omega _{d}$ is the dimensionless dark radiation energy density parameter.) More generic Weyl fluids 
were also considered \cite{GK}, \cite{LSR}-\cite{Pal}.
Depending on how this Weyl fluid evolves, 
its present day amount can be either negligible or not. It is this aspect we wish to consider here, based on our previous 
analysis \cite{supernova1}-\cite{supernova2}.  

Various brane-world models were confronted with supernova data \cite{Sahni}-\cite{Fay}, however in all these models 
the contribution of Weyl fluid was dropped, assuming it was pure 
dark radiation during all stages all the cosmological evolution. In our analysis we keep the contribution of the 
(non-radiation like) Weyl fluid. 
We already gave the analytical expression in terms of elliptical integrals for the luminosity distance when the Weyl 
contribution is small \cite{supernova1} and can be considered a perturbation. \footnote{ 
We note that for a wide class of phantom Friedmann cosmologies similar analytical results in terms of elementary and 
Weierstrass elliptic functions for the luminosity distance are available \cite{DabrowskiS}.} 
Then we have tested the models with Weyl fluid characterizing a bulk-brane energy exchange, by comparing their predictions 
with the best available supernova data \cite{supernova2}. Here we present additional analysis and strenghten our conclusions. 

Our model consists of a spatially flat FLRW brane embedded symmetrically into a 5-dimensional Vaidya-anti de Sitter bulk.  
The latter has a cosmological constant $\widetilde{\Lambda }$, black holes with masses $m$ on either sides of the brane 
and radiation. If the radiation is swithced off, $m$ is constant, and the bulk becomes Schwarzschild-anti de Sitter. 
In this configuration the Weyl fluid appears as dark radiation on the brane (with energy density $m/a^{4}$). 
Any radiation escaping from the brane causes $m$ to vary. The ansatz $m=m_{0}a^{\alpha }$, with $\alpha =2,~3$, 
comparable with structure formation has been recently advanced \cite{Pal}. 

The brane coupling constant $\kappa ^{2}$ is related to the coupling constant in the bulk $\widetilde{\kappa }^{2}$ 
and brane tension $\lambda$ as $6\kappa ^{2}=\widetilde{\kappa }^{4}\lambda $. The relation between the brane tension 
and the bulk and the brane cosmological constant $\Lambda $ is 2$\Lambda =\kappa ^{2}\lambda +\widetilde{\kappa }^{2}\widetilde{\Lambda }$.

We introduce the following dimensionless quantities: 
\begin{equation}
\Omega _{\rho } =\frac{\kappa ^{2}\rho_{0}}{3H_{0}^{2}}\ ,\ \ \ \ 
\Omega _{\lambda} =\frac{\kappa ^{2}\rho _{0}^{2}}{6\lambda H_{0}^{2}}\ , \ \ 
\Omega _{d}=\frac{2m_{0}}{a_{0}^{4-\alpha}H_{0}^{2}}\mathrm{\ },\ \ \ \\ 
\Omega _{\Lambda }=\frac{\Lambda }{3H_{0}^{2}}\ , \label{omd}
\end{equation}
where $\Omega _{tot} =\Omega_{\Lambda }+\Omega _{\rho }+\Omega _{\lambda}+\Omega_{d}$. Here $H$ is the Huble parameter, $\rho $ is the matter 
energy density on the brane and $a$ is the scale factor. The subscript $0$ denotes the present value of the 
respective quantities. The Friedmann equation written in these parameters becomes
$H^{2}/H_{0}^{2}=\Omega _{\Lambda }+\Omega _{\rho}{a_{0}^{3}}/{a^{3}}+\Omega _{d}{a_{0}^{4-\alpha }}/{a^{4-\alpha }}
+\Omega _{\lambda}{a_{0}^{6}}/{a^{6}}$ .
At present time this gives $\Omega _{tot}=1$.

\begin{figure}[!b]
 \resizebox{.7\columnwidth}{!}
  {\includegraphics[width=8cm,draft=false]{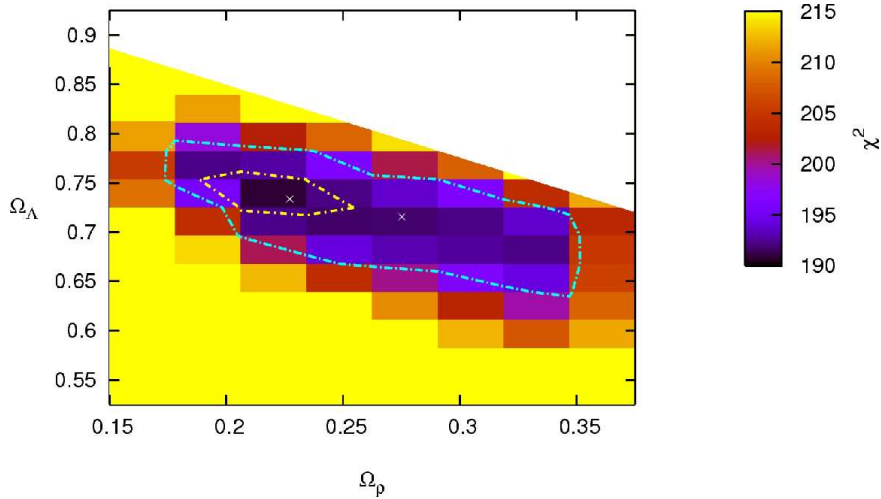}}
 \caption{For $\alpha=0$ the global minimum is at $\Omega _{\Lambda }=0.735$, $\Omega _{\rho }=0.225$ implying $\Omega_d=0.04$. The $\Lambda$CDM model
is at $\Omega _{\Lambda }=0.725$ and $\Omega _{m}=0.275$. The white area represents a forbidden parameter range.}
 \label{Fig1}
\end{figure}

\begin{figure}[!b]
 \resizebox{.7\columnwidth}{!}
  {\includegraphics[width=8cm, draft=false]{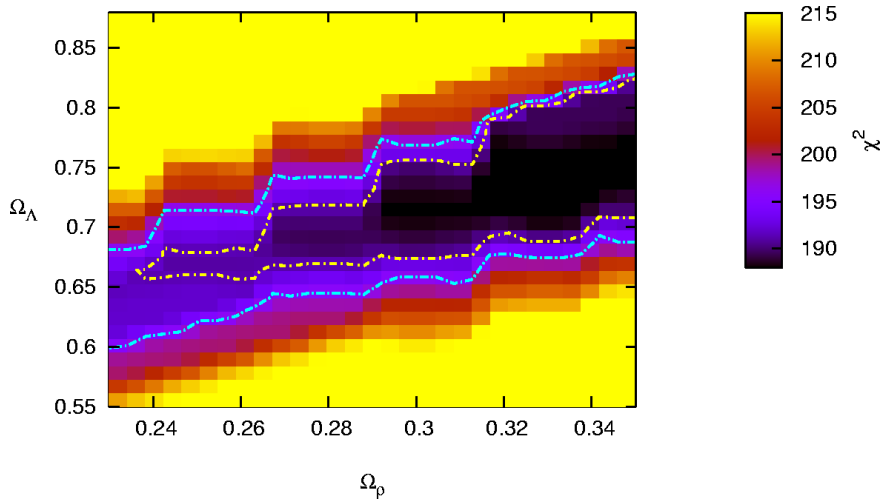}}
 \caption{For $\alpha =2$ a trough takes the place of the global minimum, showing that the compatibility of the Weyl fluid models 
with $\alpha=2$ to supernova data does not depend crucially on the particular (small) value of $\Omega_d$.}
 \label{Fig2}
\end{figure}

\begin{figure}[!b]
 \resizebox{.7\columnwidth}{!}
  {\includegraphics[width=8cm,draft=false]{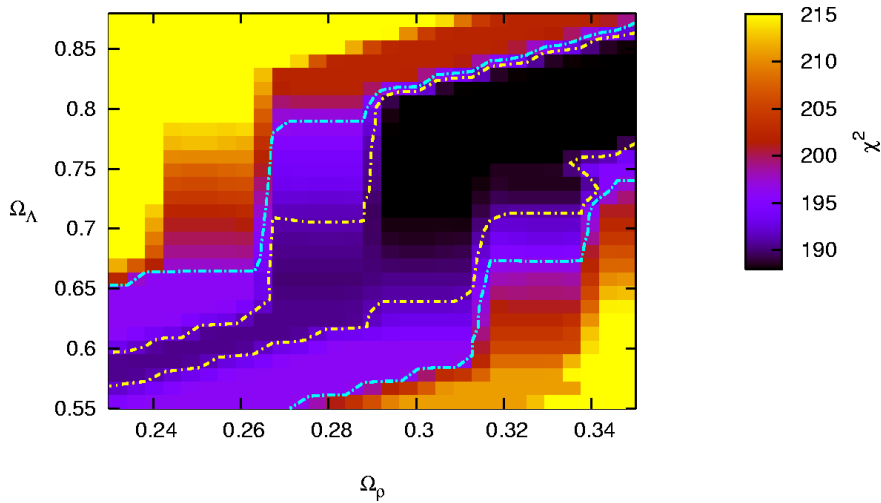}}
 \caption{Same as on Fig\ref{Fig2}, but for the $\alpha =3$ models.}
 \label{Fig3}
\end{figure}
 
The luminosity distance for the spatially flat FLRW brane becomes
\begin{equation}
d_{L}\left( z\right) =\frac{\left( 1+z\right) a_{0}}{H_{0}} \int_{a_{em}}^{a_{0}}\frac{ada}{\left[ \Omega _{\Lambda }a^{6}
+\Omega _{\rho }a_{0}^{3}a^{3}+\Omega _{d}a_{0}^{4-\alpha }a^{\alpha +2}+\Omega _{\lambda }a_{0}^{6}\right] ^{1/2}}\ .  
\label{chi2} 
\end{equation} 
The above complicated integral has no analytical form in the majority of cases. However for small $\Omega_d$ and 
(as noted earlier)) vanishing $\Omega_{\lambda}$ this integral could be given analytically as
$d_{L}^{\Lambda\lambda d}=d_{L}^{\Lambda \mathrm{CDM}}+\Omega _{d}I_{d}$,
where the coefficient $I_d$ is an analytic expression of elementary functions 
and elliptic integrals of the first and second kind \cite{supernova1}, having different forms depending on the actual 
value of the parameter $\alpha$. 

We have then compared the predictions of the models characterized by various values of $\alpha$ with the Gold 2006 supernovae data 
set \cite{gold06}, for the range $-0.1<\Omega _{d}<0.1 $ up to $z=3$.

The data selects among the brane-world models with $\alpha=0 $ a global minimum at 
$\Omega _{\Lambda }=0.735$, $\Omega _{\rho }=0.225$ and $\Omega_d=0.04$. 
We note that the value of $\Omega_{\rho}$ is in perfect agreement with
the WMAP 3-year data \cite{WMAP3y}.  
The 1-$\sigma $ and 2-$\sigma $ confidence levels
are shown of Fig\ref{Fig1} in the $\Omega _{\Lambda }-\Omega _{\rho }$ plane. The $\Lambda $CDM model is 
contained in the 2-$\sigma$ cofidence level at $\Omega _{\Lambda }=0.725,\ \Omega_{\rho}=0.275$.  
The forbidden (white) region appears because the Friedmann equation combined with 
$\Omega _{tot}=1$ gives constraints on the allowed range of $\Omega _{d}-\Omega _{\rho }$ \cite{supernova2}.

The cases $\alpha =2$ and $\alpha =3$ are represented on Fig\ref{Fig2} and Fig\ref{Fig3}, respectively.
In these cases the determination of the global minimum becomes unreliable, as the 1-$\sigma$ contours become much elongated.
Instead of a peak we have a \textquotedblleft trough\textquotedblright, which lies aslope in the 
$\Omega _{\Lambda }$-$\Omega _{\rho}$ plane. 
Together with the increase in $\alpha$ the 1-$\sigma$ contour becomes more elongated 
and turning counter-clock-wise.

This feature of the 1-$\sigma$ contours indicates that with increasing $\alpha$ the models become more likely to be compatible with the
available supernova data, irrelevant of the exact (but small) value of $\Omega_d$. This practically means that any small 
amount of Weyl fluid with $\alpha=2,~3$ is perfectly compatible with supernova data.  

In the investigated models the Weyl fluid is either dark radiation ($\alpha=0$) or describes a situation when the brane radiates into the bulk, 
feeding the bulk black holes ($\alpha=2,~3$). In the latter cases the energy density of the Weyl fluid decreases slower, than that of matter, therefore a 
sizeable Weyl fluid contribution nowasays is perfectly compatible with BBN constraints.  

For $\alpha=0 $ the Weyl fluid evolves as radiation, therefore the BBN constraints can be satisfied only with an infinitesimal amount of dark 
radiation nowadays. The preferred value of $\Omega _{d}=0.04$ would give too much dark radiation in the past. 
However if we assume that $\alpha=0$ is only a recent characteristic of the brane, this equilibrium situation being preceded by an 
epoch in which the brane is allowed to radiate, the BBN constraints can be obeyed with a small, but non-negligible amount of dark radiation nowadays.  
The BBN contraint \cite{BBN} can be satisfied \cite{supernova2}, if the brane radiates in the interval ($z_{1}$, $z_{\ast }$). 
Assuming $z_{1}=3$ the constraint for $z_{\ast }$ with different value of $\alpha$ gives $z_{\ast }\geq$     6114.20  for $\alpha =1 $,
$z_{\ast }\geq$  155.40  for $\alpha =2 $, $z_{\ast }\geq$    45.08  for $\alpha =3 $, $z_{\ast }\geq$    24.01  for $\alpha =4 $.     
Thus the known history of the Universe can be explained if the brane radiates during at least some period of the cosmological evolution. 
At early times the radiation of the brane leads to a black hole, which can further grow during structure formation. 

None of the investigated models, compatible with the available supernova data and structure formation, can be excluded by 
present observational accuracy. 
The differences among the predictions of the models are however increasing with redshift, therefore future measurements 
of very far supernovae will be able to either support or falsify these cosmological models.

\textit{Acknowledgments}: this work was supported by OTKA grants no. 46939 and 69036. L\'{A}G 
and GyMSz were further supported by the J\'{a}nos Bolyai Fellowship of the Hungarian Academy of Sciences.

\end{document}